\documentclass[prb,showpacs,floatfix,twocolumn,byrevtex,superscriptaddress]{revtex4-1}
\usepackage{amsmath}
\usepackage{graphicx}
\usepackage{float}
\usepackage{bm}
\usepackage{multirow}
\usepackage{dcolumn}

\usepackage{ulem}
\usepackage[dvipsnames,usenames]{color}

\newcommand{\icm}{\ensuremath{~\textrm{cm}^{-1}}}
\newcommand{\TMO}{TbMnO$_3$}

\newcolumntype{.}{D{.}{.}{-1}}
\begin{document}

\bibliographystyle{apsrev}

\title{Infrared reflectivity of the phonon spectra in multiferroic TbMnO$_3$}
\author{R. Schleck}
\affiliation{Laboratoire de Physique et d'\'Etude des Mat\'eriaux (LPEM) CNRS, UPMC, ESPCI-ParisTech, 10 rue Vauquelin, F-75231 Paris Cedex 5, France}

\author{R. L. Moreira}
\affiliation{Departamento de F\'{\i}sica, ICEx, Universidade Federal de Minas Gerais, CP702, 30123-970 Belo Horizonte MG, Brazil}

\author{H. Sakata}
\affiliation{Department of Physics, Tokyo University of Science, 1-3 Kagurazaka Shinjuku-ku, Tokyo 162-8601, Japan }

\author{R. P. S. M. Lobo}
\affiliation{Laboratoire de Physique et d'\'Etude des Mat\'eriaux (LPEM) CNRS, UPMC, ESPCI-ParisTech, 10 rue Vauquelin, F-75231 Paris Cedex 5, France}

\date{\today}
\begin{abstract}
We measured the temperature dependent infrared reflectivity spectra of \TMO\ with the electric field of light polarized along each of the three crystallographic axes. We analyzed the effect, on the phonon spectra, of the different phase transitions occurring in this material. We show that the antiferromagnetic transition at $T_N$ renormalizes the phonon parameters along the three directions. Our data indicate that the electromagnon, observed along the $a$ direction, has an important contribution to the building of the dielectric constant. Only one phonon, observed along the $c$-axis, has anomalies at the ferroelectric transition. This phonon is built mostly from Mn vibrations, suggesting that Mn displacements are closely related to the formation of the ferroelectric order.
\end{abstract}
\maketitle

%
%
\section{Introduction}
The growing interest in multiferroics was triggered by their potential industrial applications.\cite{Eerenstein2006} The current research focus is the puzzling mechanism driving the coupling between the magnetic and the ferroelectric (FE) orders, simultaneously present in these materials.\cite{Katsura2005,Hu2008,Sergienko2006a,Aguilar2009} Indeed, in the so-called improper multiferroics, the ferroelectric polarization is a consequence of a particular magnetic order which breaks the inversion symmetry of the crystal.\cite{Kenzelmann2005} In this class of materials, the magnetic origin of the ferroelectricity is responsible for a strong coupling between both orders which, for example, makes it possible to rotate the direction of the spontaneous electric polarization by an applied magnetic field.\cite{Kimura2003a}

\TMO\ belongs to this class of improper multiferroics. It is paramagnetic and paraelectric at high temperatures. Below $T_N\approx 41$~K, the Mn spins order antiferromagnetically almost\cite{Wilkins2009} parallel to the $b$-axis with an incommensurate wavevector $\mathbf{k}\approx(0,0.28,1)$ (Ref.~\onlinecite{Kenzelmann2005}). At $T_C \approx 28$~K the Mn spins acquire a $c$-axis component which transforms the sinusoid into a cycloid magnetic order with approximately the same wavevector. This cycloid of spins breaks the inversion symmetry of the crystal,\cite{Kenzelmann2005} and induces a spontaneous electric polarization along the $c$-axis.\cite{Kimura2003a} In addition to the fact that the cycloid magnetic and ferroelectric transitions occur at the same temperature, the magnetic origin of the ferroelectric order was evidenced by the fact that the electric polarization is tied to the cycloid plane\cite{Aliouane2009} and chirality.\cite{Yamasaki2007} The electric polarization is given by $\mathbf{P}=\mathbf{e}_{i,i+1}\times(\mathbf{S}_i\times\mathbf{S}_{i+1})$ where $\mathbf{S}_i$ and $\mathbf{S}_{i+1}$ are the spins on sites $i$ and $i+1$ and $\mathbf{e}_{i,i+1}$ is the unit vector directed from site $i$ to site $i+1$. This magnetic spiral induced ferroelectricity was theoretically explained by Katsura, Nagaosa and Balatsky\cite{Katsura2005} as an inverse Dzyaloshinsky-Moriya interaction (IDM) between noncollinear spins creating an electronic polarization. Inelastic neutron scattering could not detect any $c$-axis phonon anomaly related to ferroelectricity, suggesting an electronic origin for the ferroelectricity.\cite{Kajimoto2009} An alternative scenario is to consider that the spiral magnetic order induces ionic displacements producing the ferroelectric polarization. This mechanism was proposed by \textcite{Malashevich2008} whose \textit{ab-initio} calculations showed that in the $ac$-spiral state of \TMO\ the ferroelectric polarization was mainly due to ionic displacements in the lattice. 

In \TMO, there is also a strong dynamic ma\-gnetoelectric coupling which results in the existence of electric dipole active magnetic excitations called electro\-magnons.\cite{Pimenov2006a} This electromagnon was measured by inelastic neutron scattering,\cite{Senff2007,Senff2008a} Raman,\cite{Rovillain2010} infrared (IR)\cite{Takahashi2008} and terahertz\cite{Pimenov2006a} spectroscopies and was found to be excited by an electric field parallel to the $a$-axis regardless of the cycloid plane orientation.\cite{Aguilar2009,Pimenov2009} A theory based on Heisenberg exchange was proposed by \textcite{Aguilar2009} to explain this selection rule for the low frequency excitations. These observations suggest that the electromagnon is tied to the lattice rather than to the spin cycloid. Moreover, \textcite{Takahashi2008} evidenced a coupling between the electromagnon and the lower energy phonon in the $a$-axis direction. For these reasons, we can expect to get useful information about the magneto-electric coupling in \TMO\ through the study of its lattice dynamics. 

Previous IR measurements on \TMO, by Schmidt et al.,\cite{Schmidt2009} were performed along the $a$ and $c$ axis, but not the $b$ direction. Although they have some uncertainties in phonon assignments due to polarization leakage, they found that the $a$ and $c$ phonon frequencies change at $T_N$. In order to investigate more thoroughly the phonon behavior of TbMnO$_3$ around the antiferromagnetic (AFM) and FE transitions we have undertaken a detailed temperature dependent infrared study of the phonon spectra of  \TMO\ with the electric field of light polarized along each of the three orthorhombic directions. We were able to unambiguously identify the infrared active normal modes predicted by group theory. The analysis of the thermal evolution of the mode parameters shows that the AFM transition affects the phonon spectra equally in all directions. In other words, the phonon renormalization at $T_N$ along the directions ($a$ and $c$) that dominate the dielectric properties are of the same size as the changes observed along the magnetic order $b$ axis. We also find that along the polarization axis ($c$ direction) additional changes appear at the ferroelectric phase transition. This is the first evidence in \TMO\ of a phonon anomaly related to the ferroelectric order, adding to the debate on whether or not the lattice plays any role in the magnetoelectric coupling.

%
%
\section{Experimental}
The samples used in this study were grown by the floating zone technique. They were aligned using Laue X-ray back reflection and showed dielectric anomalies at 42 K and 27 K corresponding to $T_N$ and $T_C$, respectively.\cite{Sakata2007} 

We measured the near normal (10$^\circ$) incidence reflectivity of two \TMO\ samples with their largest face cut, respectively, perpendicular to the $b$-axis ($ac$ face) and perpendicular to the $a$-axis ($bc$ face). These samples allowed us to measure the reflectivity with the electric field of light polarized along each of the 3 orthorhombic directions at 26 different temperatures between 5~K and 300~K in the spectral range [40--6000]\icm. In order to obtain the absolute reflectivity of the sample, we used an \textit{in situ} gold overfilling technique.\cite{Homes1993} With this technique, we can achieve an absolute accuracy in the reflectivity better than 1\%, and the relative error between different temperatures is of the order of 0.1\%. The data were collected in an Bruker IFS66/s Fourier transform spectrometer, the sample being attached to the cold finger of an ARS Helitran cryostat.

%
%
\section{Results}
\TMO\ crystallizes in an orthorhombically distorted perovskite structure with space group $Pbnm$ ($D_{2h}^{16}$). The irreducible representations decomposition for the normal modes in this system is $7A_g \oplus 8A_u \oplus 5B_{1g} \oplus 10B_{1u} \oplus 7B_{2g} \oplus 8B_{2u} \oplus 5B_{3g} \oplus 10B_{3u}$ among which $9 B_{1u}$, $9 B_{3u}$ and $7 B_{2u}$ are IR active for the electric field of light $\mathbf{E}$ parallel to the $a$, $b$ and $c$ axes, respectively. The $7 A_g$, $5 B_{1g}$, $7 B_{2g}$ and $5 B_{3g}$ are Raman active, all the remaining $8 A_u$ modes are silent.
\begin{figure}[htb]
  \includegraphics[width=8cm]{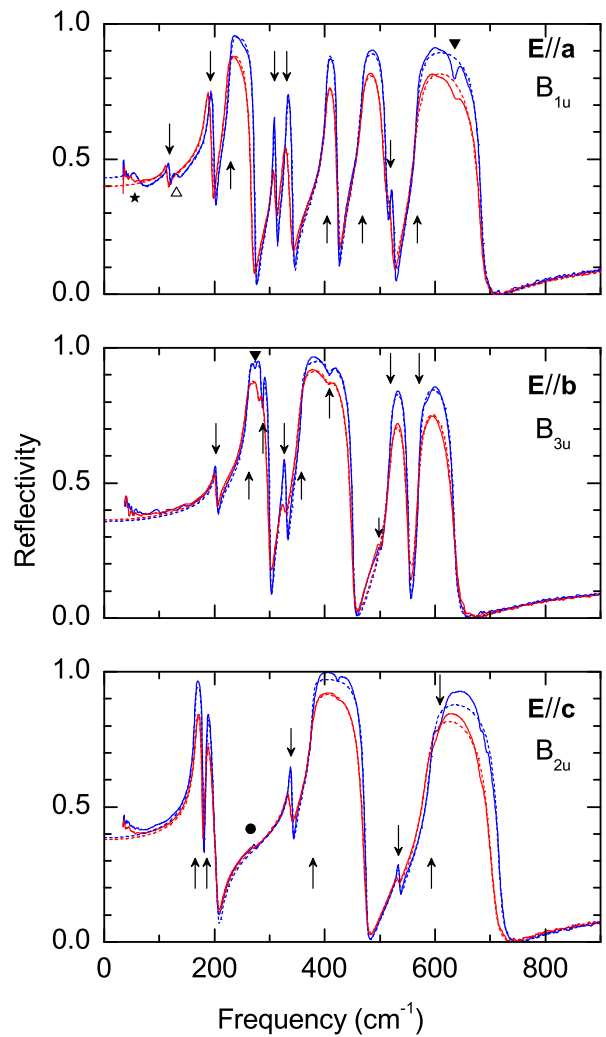} 
  \caption{(color online) Reflectivity of \TMO\ measured with the electric field of light ($\mathbf{E}$) polarized along each of the three crystallographic axis. Solid lines are the measured reflectivity at 5~K (blue) and 300~K (red), and the dotted lines are fits to the data with Lorentz oscillators. The arrows represent the positions of the TO phonon branches. As discussed in the text, the star in the top panel indicates the electromagnon response and the open up-triangle a possible  Tb cristal field transition. In the top and middle panels, the solid down-triangles indicate LO phonon leakage due to a non normal incidence measurement. In the bottom panel, the solid circle indicates a phonon leakage due to a slight in-plane misalignement.}
  \label{reflectivity}
\end{figure}

The solid lines on Fig.~\ref{reflectivity} are the reflectivity spectra for the three polarizations measured at 5~K and 300~K. The top panel shows the reflectivity measured with $\mathbf{E}\parallel\mathbf{a}$. We find the $9 B_{1u}$ modes predicted by group theory, which transverse optical (TO) frequencies are indicated by the arrows. Some features in these spectra deserve further discussion. Indicated by a solid down-triangle, at 635\icm\ there is a dip in the higher frequency phonon peak. This is not an additional mode of $B_{1u}$ symmetry. It is rather a signature of the highest $B_{3u}$ ($\mathbf{E}\parallel \mathbf{b}$) longitudinal optical (LO) branch due to the fact that the incident light is not exactly normal to the sample's surface. This measurement was done with a $p$ polarized light, which leads to the excitation of longitudinal modes pertaining to the crystallographic direction normal to the surface (the $b$-axis in the present case).\cite{Duarte1987} The peak around 130\icm, indicated by the open up-triangle, appears at temperatures below 150~K and was already observed in reflection\cite{Schmidt2009} and transmission\cite{Takahashi2008} measurements. It was tentatively assigned to the excitation of a double zone-edge magnon\cite{Takahashi2008}, a one magnon+one phonon mode\cite{Schmidt2009} or a transition between two crystal field split levels of the Tb ions. At lower wavenumbers, there is a considerable evolution of the spectra below 40 K with the apparition of an electromagnon peak at 60\icm, indicated by the star. A detailed study of the electromagnon is the subject of a subsequent paper.

The middle panel of Fig.~\ref{reflectivity} shows the reflectivity measured with $\mathbf{E}\parallel\mathbf{b}$. In this direction, group theory predicts 9 B$_{3u}$ infrared active modes, which TO frequencies are indicated by the arrows. The dip at 273\icm\ in the second phonon peak marked by the solid down-triangle is a signature, once more due to the non normal incidence measurement, of the sharp LO branch of the third phonon in the $a$ direction ($B_{1u}$ symmetry).

On the bottom panel of Fig.~\ref{reflectivity}, we show the $c$-axis reflectivity spectra at low and high temperatures. From a group theory analysis, we expect 7 B$_{2u}$ IR active modes, all of them are visible on the reflection spectrum as marked by arrows. The structure indicated by the solid circle is a polarization leakage from the $a$ direction due to an in-plane misorientation of the sample, which we estimate to be less than 2\%. The increase of the reflectivity below 75\icm\ is an additional reflection on the back surface of the sample which becomes transparent in this region.

In order to quantitavely analyze the temperature evolution of the spectra, we modeled the data using one Lorentz oscillator for each phonon mode. In this model, the dielectric funtion is given by :
%
\begin{equation}
\varepsilon(\omega)=\varepsilon_\infty+\sum_k\frac{\Delta\varepsilon_k \Omega_{TO_k}^2}{\Omega_{TO_k}^2-\omega^2-i\gamma_k\omega},
\label{lorentz}
\end{equation}
%
where $\varepsilon_\infty$ is the contribution from electronic transitions to the dielectric function, and each phonon is described by a resonance frequency $\Omega_{TO_k}$, an oscillator strength $\Delta\varepsilon_k$, and damping $\gamma_k$. These parameters were fitted so that the reflectivity at normal incidence, given by $R = |1 - \sqrt{\varepsilon}|^2 / |1 + \sqrt{\varepsilon}|^2$, matches the experimental data. Fits are shown as dotted lines on Fig.~\ref{reflectivity}. 
 
The Lorentz model fits were complemented by Kramers-Kronig analysis. This allowed us to ascertain that very small changes observed in the fitting parameters can be tracked directly to the experimental dielectric function.

%
%
\section{Discussion}

The parameters of all of the IR-active phonon modes at low (5~K) and high (300~K) temperatures are shown in Table~\ref{table_params}. Except for the two lower frequency $B_{2u}$ $c$-axis phonons, all eigenfrequencies are higher at low temperature. This is the usual behavior for phonons which tend to soften due to the dilatation of the unit cell when the temperature is raised. This table also shows the oscillator strengths of each phonon as well as the contribution of phonons and higher energy excitations to the static dielectric constant. The longitudinal frequencies shown are calculated as the zeros of the dielectric function with all the $\gamma$'s set to zero.

\begin{figure}[h!tb]
  \includegraphics[width=8cm]{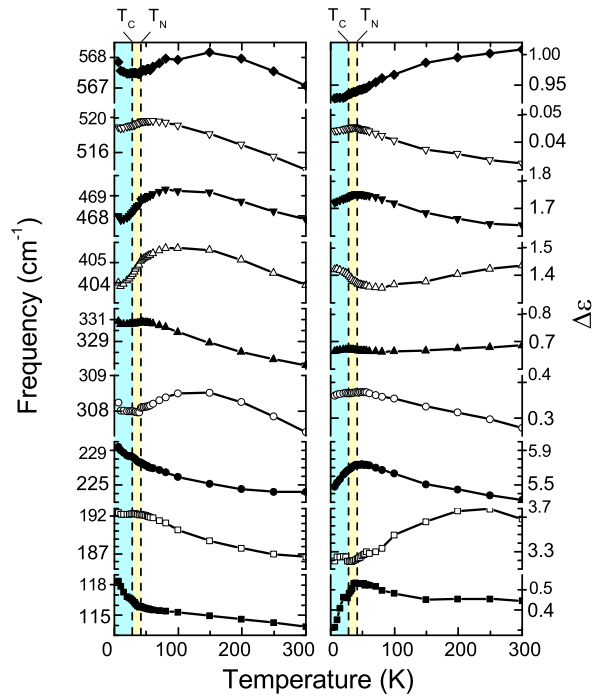} 
  \caption{Thermal evolution of the phonon eigenfrequencies and oscillator strengths for $\mathbf{E} \parallel \mathbf{a}$ $(B_{1u})$ between 5~K and 300~K. The phase transition temperatures $T_N$ and $T_C$ are indicated by the dotted lines.}
  \label{params_a}
\end{figure}
In order to analyze the details of the effects of the phase transitions on the phonon spectrum in each direction, we followed the evolution of the frequencies and strengths of each phonon with the temperature. We will not discuss $\gamma$ values, as they not have any anomaly at $T_C$ or $T_N$, within our fitting accuracy.

Figure~\ref{params_a} shows the thermal evolution of the fitting parameters for $\mathbf{E} \parallel \mathbf{a}$. In this direction we observe similar changes as those seen by \textcite{Schmidt2009} There is a clear renormalization of the phonon parameters at $T_N$ and within experimental accuracy, we see no changes at $T_C$. The lowest frequency phonon is the most affected one as its eigenfrequency is increased by 3\% below $T_N$. Even more striking, its oscillator strength is reduced by almost 50\% between $T_N$ and 5 K. These changes are partly due to static spin-phonon coupling, like those occurring at $T_N$ in MnF$_2$,\cite{Schleck2009} but also to the coupling of the phonon spectrum with the electromagnon developing in the AFM phases.\cite{Takahashi2008} It is also worth mentioning that \textcite{Laverdiere2006} did not detect any spin-phonon coupling by Raman spectroscopy in \TMO.

\begin{figure}[htb]
  \includegraphics[width=8cm]{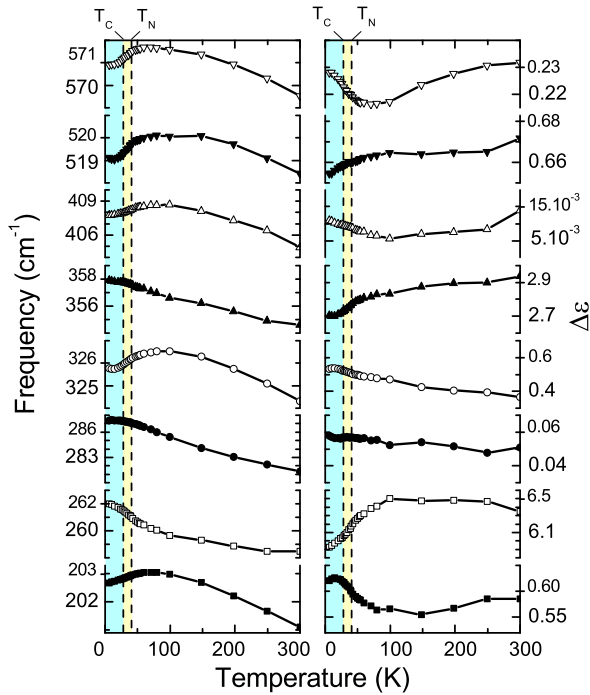} 
  \caption{Thermal evolution of the phonon eigenfrequencies and oscillator strengths for $\mathbf{E}\parallel \mathbf{b}$ $(B_{3u})$ between 5~K and 300~K. The phase transition temperatures $T_N$ and $T_C$ are indicated by the dotted lines. A phonon around 500~\icm\ is not shown as, due to its very small oscillator strength, it could not be fitted accurately over the whole temperature range.}
  \label{params_b}
\end{figure}
The phonon parameters along the $b$ direction, shown on Fig.~\ref{params_b}, are slightly less influenced by the AFM order but they still exhibit clear changes at $T_N$. The eigenfrequencies continuously increase from room temperature down to 100~K. Below this temperature the eigenfrequency of each phonon mode is either further increased or decreased down to 5 K, surprisingly showing very little or no saturation. The physical response of \TMO\ along the $b$ axis is dominated by the magnetic ordering rather than the system dielectric properties. Hence, the spectra renormalization observed along this direction is mostly due to spin-phonon coupling.\cite{Schleck2009} 

\begin{figure}[htb]
  \includegraphics[width=8cm]{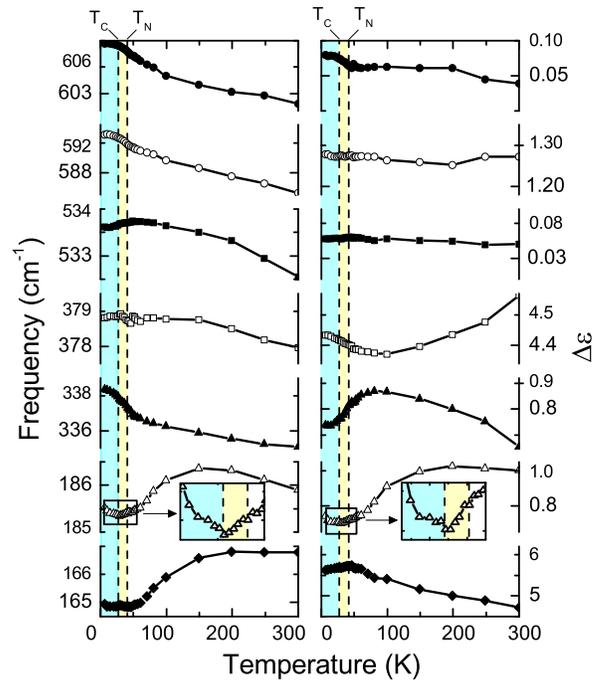} 
  \caption{Thermal evolution of the phonon eigenfrequencies and oscillator strengths for $\mathbf{E}_{light}\parallel \mathbf{c}$ $(B_{2u})$ between 5~K and 300~K. The phase transition temperatures $T_N$ and $T_C$ are indicated by the dotted lines. Insets are a zoom at low temperatures for the $B_{2u}(2)$ mode, which shows the largest change at $T_C$.}
  \label{params_c}
\end{figure}
Figure~\ref{params_c} shows the parameters for the $c$ direction (static electric polarization axis) phonons. Although the most significant dielectric signatures (polarization and electromagnons) of \TMO\ show up in the $c$ and $a$ directions, the phonon renormalization along these axes are of the same order of magnitude as those along the magnetic order $b$ axis. Therefore, we can conclude that these changes are mostly due to the spin-phonon coupling.

However our data also shows that the ferroelectric order has a clear signature along the $c$ direction. This is particularly visible on the 185\icm\ phonon mode softening when approaching $T_C$ from high and low temperatures. As the spontaneous polarization develops along this axis, it is worth taking a closer look at the $B_{2u}$ normal modes. 

\begin{figure}[htb]
  \includegraphics[width=8cm]{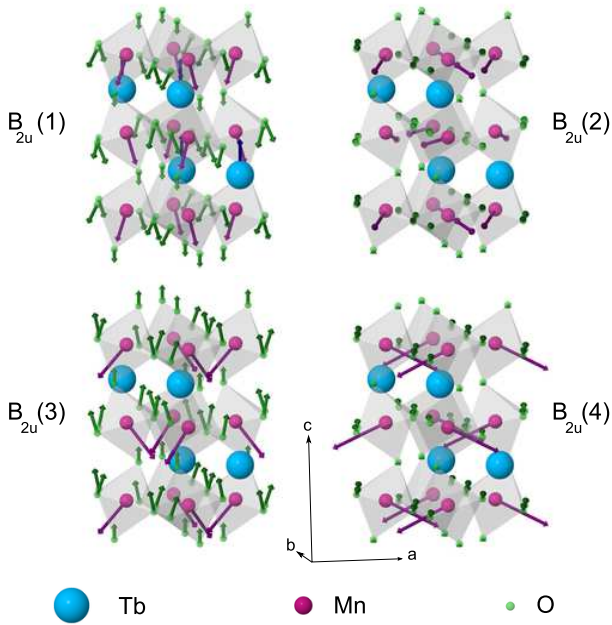} 
  \caption{(color online) Normal modes of $B_{2u}$ symmetry involving displacements of Mn atoms.}
  \label{modes_c}
\end{figure}
Figure~\ref{modes_c} shows the atomic displacements for the four lowest frequency $B_{2u}$ modes. These are the four modes that involve displacements of Mn atoms for this polarization. Modes 3 and 4 are the ones that show the strongest changes in their oscillator strengths below $T_N$. Indeed, the atomic displacements for these modes are the ones that change the Mn-O bonds the most, as the $c$ component of Mn and O movements are in opposite directions. As a consequence, they contribute more to the renormalization of the AFM exchange ($J$), in a process similar to the one observed in MnF$_2$.\cite{Schleck2009} Mode $B_{2u}(1)$ involves Mn and O displacements in the same direction. As a result, this phonon does not contribute to a large renormalization of either $J$ or the unit cell charge distribution. This is in agreement with the small parameters changes seen for this phonon at $T_N$ and $T_C$. 

Mode $B_{2u}(2)$ shows a most interesting behavior. It has a clear anomaly in both oscillator strength and frequency at $T_C$ (see insets in Fig.~\ref{params_c}), indicating that Mn displacements are closely linked to the formation of the FE order. In the framework of the IDM mechanism involving ionic displacements, we expect a distortion of the MnO$_6$ octaedra with the appearance of the FE polarization.\cite{Sergienko2006a} However some questions remain on whether the important displacements related to the FE order are in the O (Ref.~\onlinecite{Sergienko2006a}) or Mn (Ref.~\onlinecite{Xiang2008}) atoms. As the  $B_{2u}(2)$ phonon involves mostly displacements of Mn (the oxygen atoms oscillation amplitude is the smallest of the four modes shown), the anomalies observed at $T_C$ support the latter picture.

\begin{table*}
\begin{center}
\caption{Lorentz fit parameters at 5~K and 300~K. Units for $\Omega_{TO}$, $\Omega_{LO}$ and $\gamma$ are \icm.  Fitted values for $\varepsilon_\infty$ are 5.73 along the $a$-axis 5.07 along the $b$-axis and 5.37 along the $c$-axis. $B_{1u}$, $B_{2u}$ and $B_{3u}$ modes are active for the electric field of light parallel to, respectively, $a$, $c$ and $b$ axis. The longitudinal frequencies ($\Omega_{LO}$) are not input parameters to the Lorentz model. They were calculated by finding the zeros of the real part of the dielectric function in the absence of damping ($\gamma = 0$ for all phonons).
\label{table_params}}
\begin{ruledtabular}
\begin{tabular}{c.............}
\multicolumn{1}{c}{}	&	\multicolumn{4}{c}{5K}& & \multicolumn{4}{c}{300K} \\
Mode	& \multicolumn{1}{c}{$\Omega_{TO}$ }& \multicolumn{1}{c}{$\Omega_{LO}$ }&	\multicolumn{1}{c}{$\Delta\epsilon$} &\multicolumn{1}{c}{ $\gamma$ }& &\multicolumn{1}{c}{$\Omega_{TO}$ }& \multicolumn{1}{c}{$\Omega_{LO}$ }&	\multicolumn{1}{c}{$\Delta\epsilon$} &\multicolumn{1}{c}{ $\gamma$ } \\
\hline
$B_{1u}$ & 118.4 & 119.1 & 0.31 &  3.50 && 113.8 & 114.9 & 0.45 &  4.91 \\
$B_{1u}$ & 192.4 & 200.9 & 3.19 &  5.81 && 186.6 & 196.2 & 3.68 &  6.70 \\
$B_{1u}$ & 229.4 & 274.4 & 5.48 &  3.10 && 224.2 & 269.5 & 5.33 &  8.05 \\
$B_{1u}$ & 308.2 & 313.0 & 0.36 &  2.91 && 307.4 & 310.7 & 0.27 &  5.53 \\
$B_{1u}$ & 330.8 & 343.5 & 0.67 &  4.72 && 326.9 & 339.5 & 0.69 & 10.50 \\
$B_{1u}$ & 404.0 & 425.2 & 1.42 &  3.59 && 403.9 & 425.1 & 1.44 &  7.96 \\
$B_{1u}$ & 468.0 & 514.1 & 1.72 &  7.46 && 467.8 & 510.6 & 1.64 & 12.54 \\
$B_{1u}$ & 519.0 & 526.8 & 0.04 &  7.51 && 514.0 & 521.7 & 0.03 & 10.45 \\
$B_{1u}$ & 567.8 & 683.5 & 0.93 & 10.82 && 567.1 & 683.1 & 1.01 & 19.47 \\
\multicolumn{3}{l}{$\varepsilon_\infty+\sum{\Delta\varepsilon}$}& 19.85 & & & & &20.27 & \\
\\ 
$B_{2u}$ & 165.0 & 180.7 & 5.63 &  1.53 && 166.8 & 179.8 & 4.72 &  5.20 \\
$B_{2u}$ & 185.5 & 203.6 & 0.74 &  3.26 && 185.9 & 204.0 & 1.00 &  5.76 \\
$B_{2u}$ & 338.3 & 342.5 & 0.74 &  4.82 && 335.1 & 338.8 & 0.65 &  9.12 \\
$B_{2u}$ & 378.8 & 475.0 & 4.43 &  2.88 && 378.0 & 474.5 & 4.56 &  9.71 \\
$B_{2u}$ & 534.2 & 536.4 & 0.06 &  6.20 && 532.1 & 533.9 & 0.05 & 14.06 \\
$B_{2u}$ & 593.1 & 607.7 & 1.28 & 12.34 && 585.2 & 601.2 & 1.27 & 19.78 \\
$B_{2u}$ & 608.6 & 716.3 & 0.08 &  9.94 && 601.8 & 705.2 & 0.04 &  8.87 \\
\multicolumn{3}{l}{$\varepsilon_\infty+\sum{\Delta\varepsilon}$}&  18.33 & & & & &17.66 & \\
\\
$B_{3u}$ & 202.7 & 205.0 & 0.62 & 4.66 && 201.1 & 203.2 & 0.59 &  5.69 \\
$B_{3u}$ & 262.0 & 286.8 & 5.93 & 3.30 && 258.5 & 281.0 & 6.35 &  8.90 \\
$B_{3u}$ & 287.4 & 301.4 & 0.06 & 2.72 && 281.3 & 299.8 & 0.05 &  3.84 \\
$B_{3u}$ & 325.8 & 331.5 & 0.53 & 5.66 && 324.3 & 328.0 & 0.36 & 12.20 \\
$B_{3u}$ & 357.9 & 407.4 & 2.71 & 5.13 && 354.6 & 404.5 & 2.94 &  9.14 \\
$B_{3u}$ & 407.7 & 451.8 & 0.01 & 7.91 && 404.9 & 451.3 & 0.01 & 11.28 \\
$B_{3u}$ & 499.0 & 499.5 & 0.02 & 4.99 && 497.7 & 499.0 & 0.04 &  7.90 \\
$B_{3u}$ & 519.1 & 553.5 & 0.66 & 6.83 && 518.4 & 552.7 & 0.67 & 13.24 \\
$B_{3u}$ & 570.9 & 634.3 & 0.23 & 7.89 && 569.5 & 635.9 & 0.23 & 13.92 \\
\multicolumn{3}{l}{$\varepsilon_\infty+\sum{\Delta\varepsilon}$}& 15.91 & & & & &16.27 & \\
\end{tabular}
\end{ruledtabular}
\end{center}
\end{table*}
We can also compare the measured static dielectric constant $\varepsilon_0$ to the DC limit of the dielectric function extrapolated from the IR spectra. The contribution of the phonon and electronic excitations calculated from the fit to our data is given by $\varepsilon_{\it LF} = \varepsilon_{\infty} + \sum{\Delta\varepsilon}$. The static dielectric constant was measured by \textcite{Kimura2003a} at 10~kHz. The zero frequency infrared extrapolation gives, at $T = 5$~K (see Table~\ref{table_params}), $\varepsilon_{\it LF}^{(a)} \approx 20$; $\varepsilon_{\it LF}^{(b)} \approx 16$ and $\varepsilon_{\it LF}^{(c)} \approx 18$. The values obtained by \citeauthor{Kimura2003a} are $\varepsilon^{(a)} \approx 24$; $\varepsilon^{(b)} \approx 23$ and $\varepsilon^{(c)} \approx 29$. In all cases the dc value is higher than the optical one. Neither the experimental errors nor sample inhomogeneities can account for such differences which suggests the existence of lower frequency excitations. Actually, a dielectric dispersion was observed in \TMO\ (and in many other RMnO$_3$  compounds) along the $c$-axis that shows a relaxor-like behavior.\cite{Goto2004} Along \TMO\ $c$-axis, this low frequency relaxation produces a drop of 10 points in the dielectric function when the measurement frequency increases from 1 to 100~kHz. This drop is enough to reconcile dc and IR dielectric responses. Although \textcite{Goto2004} did not observe any dispersion in the $a$ or $b$ dielectric constants at radiofrequencies, we believe that the relaxation could lie in the MHz or GHz regions. Dispersions in different crystallographic directions could have the same origin, once charge carriers usually have different activation energies in anisotropic systems. Even though our data alone cannot give a definite support the mentioned mechanism, it does require the existence of (sub)millimeter wave excitations. 

The thermal evolution of the low frequency dielectric functions along each direction can be tracked from the fitting parameters. Although the static values change along all three directions at low temperature,\cite{Kimura2003a,Goto2004} only the $a$ axis data can be compared to our infrared measurements. The changes seen along $b$ are smaller than our error bars and those along $c$ are confined to a very narrow temperature range of 0.5~K, much smaller than the temperature resolution of this work.

\begin{figure}[htb]
  \includegraphics[width=8cm]{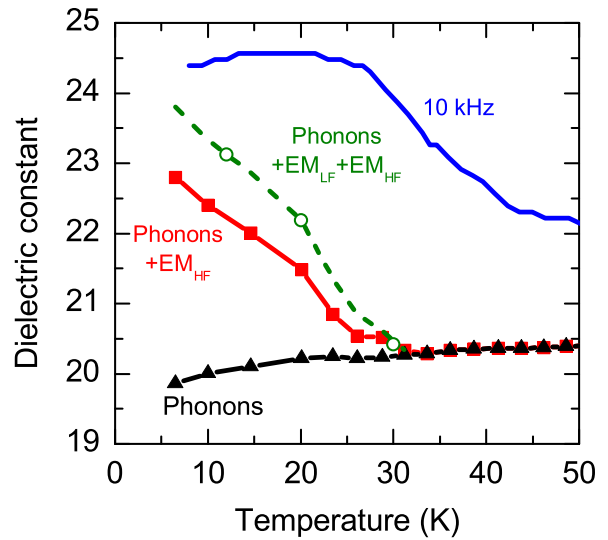} 
  \caption{(color online) The symbols are the low frequency dielectric constant calculated from the fit results for $\mathbf{E} \parallel \mathbf{a}$ between 5~K and 50~K with (solid red squares) and without (solid black triangles) the contribution of the 60~\icm\ (HF) electromagnon. The dashed green line adds the estimated contribution of the 26~\icm\ (LF) electromagnon measured by \protect\textcite{Pimenov2006a} The open cirlces are the the temperatures measured by the authors. The solid blue line is the low frequency dielectric constant at 10kHz measured by \protect\textcite{Kimura2003a}}
  \label{eps0_a}
\end{figure}
Nevertheless, from the infrared point of view, the $a$ axis dielectric constant is the most interesting as it is the direction where the electromagnon appears. Figure~\ref{eps0_a} shows the static (10~kHz) dielectric constant measured in the $a$ direction by \citeauthor{Kimura2003a} as well as the low frequency values extrapolated from our data. The triangles are the phonon contribution ($\varepsilon_{\it ph}$) and the squares include the 60 \icm\ electromagnon (which contribution $\Delta\varepsilon_{{\it EM}_{\it HF}}$ was estimated by fitting it to a Lorentz oscillator in the reflectivity spectra). At the lowest temperature, we obtained $\Delta\varepsilon_{{\it EM}_{\it HF}}\approx 3$, which gives a total of $\varepsilon_{{\it ph+EM}_{HF}}\approx 23$. \textcite{Pimenov2006a} determined that a lower frequency electromagnon exists at 26~\icm, below our measured spectral range. We calculated the oscillator strength ($\Delta\varepsilon_{{\it EM}_{\it LF}}$) for this low frequency electromagnon from \citeauthor{Pimenov2006a} data and added it to our measurements. The result is show as the dashed line in Fig.~\ref{eps0_a}, where the open circles are the temperatures measured in Ref.~\onlinecite{Pimenov2006a}. At the lowest temperature the contribution from both electromagnons solve the difference between static and (phonononic) dynamic dielectric constant values. Therefore, Fig.~\ref{eps0_a} shows that the dielectric response is mostly due to the phonons, and the magnitude of the $\varepsilon_0$ jump below $T_N$ is the same as the magnitude of the electromagnons contributions. Hence, this jump can be undoubtedly assigned to the appearance of electromagnons.

A remarkable difference occur, however, in the thermal evolution of the static and infrared constants between 6 K and 50 K. Indeed, the dielectric constant at 10 kHz is always larger than the infrared extrapolated values, including the electromagnon contribution. This result is likely related to the observation of a dielectric dispersion in the $c$ direction at radiofrequencies mentioned above. This effect, in \TMO\ and other related manganites, was tentatively attributed to the dynamics of localized carriers coupled to the lattice, which would affect the magnetic ordering.\cite{Goto2004} Moreover, since the dielectric dispersion exists even above $T_N$, one could think that fluctuations of the spin modulation wave could generate local regions with fluctuating net polarization, which average vanishes above $T_N$. This would imply a phase coexistence in a large temperature interval, as observed by \textcite{Barath2008} under applied magnetic field. Nonetheless, a definite explanation for the observed effect deserves further investigations.

%
%
\section{Conclusions}
We presented a detailed analysis of the phonon spectrum of \TMO\ and its evolution with temperature. We showed that all the modes predicted by group theory are clearly identifiable in the reflectivity spectrum of each orthorhombic direction. The analysis of the thermal evolution of the phonon parameters shows a clear coupling between the phonons in the three orthorhombic axes to the magnetic order. The phonon spectra renormalization at $T_N$ is of the same magnitude for all crystallographic directions. An important result of this study is that ferroelectricity produces an additional renormalization of the $B_{2u}(2)$ phonon along the $c$ direction. This phonon is composed of Mn vibrations only, indicating that this is the important atomic displacement induced by the IDM interaction. We evidenced that in both $a$ and $b$ directions, the static dielectric constant is mostly due to the phonons with an additional jump at $T_N$ due to the electromagnon for $\mathbf{E} \parallel \mathbf{a}$. In both cases, small sub-millimeter wave excitations are necessary to bring the dielectric function to its dc value. Along the $c$-axis the relaxor-like behavior seen in Ref.~\onlinecite{Goto2004} explains the much larger discrepancies between dc and infrared values.

\section*{Acknowledgment}
This work was partially funded through CNRS PICS program 4905. RLM acknowledges support from the Brazilian agencies FAPEMIG and CNPq.

%
%
\bibliography{biblio}

\begin{thebibliography}{27}
\expandafter\ifx\csname natexlab\endcsname\relax\def\natexlab#1{#1}\fi
\expandafter\ifx\csname bibnamefont\endcsname\relax
  \def\bibnamefont#1{#1}\fi
\expandafter\ifx\csname bibfnamefont\endcsname\relax
  \def\bibfnamefont#1{#1}\fi
\expandafter\ifx\csname citenamefont\endcsname\relax
  \def\citenamefont#1{#1}\fi
\expandafter\ifx\csname url\endcsname\relax
  \def\url#1{\texttt{#1}}\fi
\expandafter\ifx\csname urlprefix\endcsname\relax\def\urlprefix{URL }\fi
\providecommand{\bibinfo}[2]{#2}
\providecommand{\eprint}[2][]{\url{#2}}

\bibitem[{\citenamefont{Eerenstein et~al.}(2006)\citenamefont{Eerenstein,
  Mathur, and Scott}}]{Eerenstein2006}
\bibinfo{author}{\bibfnamefont{W.}~\bibnamefont{Eerenstein}},
  \bibinfo{author}{\bibfnamefont{N.~D.} \bibnamefont{Mathur}},
  \bibnamefont{and} \bibinfo{author}{\bibfnamefont{J.~F.} \bibnamefont{Scott}},
  \bibinfo{journal}{Nature} \textbf{\bibinfo{volume}{442}},
  \bibinfo{pages}{759} (\bibinfo{year}{2006}).

\bibitem[{\citenamefont{Katsura et~al.}(2005)\citenamefont{Katsura, Nagaosa,
  and Balatsky}}]{Katsura2005}
\bibinfo{author}{\bibfnamefont{H.}~\bibnamefont{Katsura}},
  \bibinfo{author}{\bibfnamefont{N.}~\bibnamefont{Nagaosa}}, \bibnamefont{and}
  \bibinfo{author}{\bibfnamefont{A.~V.} \bibnamefont{Balatsky}},
  \bibinfo{journal}{Phys. Rev. Lett.} \textbf{\bibinfo{volume}{95}},
  \bibinfo{pages}{057205} (\bibinfo{year}{2005}).

\bibitem[{\citenamefont{Hu}(2008)}]{Hu2008}
\bibinfo{author}{\bibfnamefont{J.}~\bibnamefont{Hu}}, \bibinfo{journal}{Phys.
  Rev. Lett.} \textbf{\bibinfo{volume}{100}}, \bibinfo{pages}{077202}
  (\bibinfo{year}{2008}).

\bibitem[{\citenamefont{Sergienko and Dagotto}(2006)}]{Sergienko2006a}
\bibinfo{author}{\bibfnamefont{I.~A.} \bibnamefont{Sergienko}}
  \bibnamefont{and} \bibinfo{author}{\bibfnamefont{E.}~\bibnamefont{Dagotto}},
  \bibinfo{journal}{Phys. Rev. B} \textbf{\bibinfo{volume}{73}},
  \bibinfo{pages}{094434} (\bibinfo{year}{2006}).

\bibitem[{\citenamefont{Vald\'{e}s-Aguilar
  et~al.}(2009)\citenamefont{Vald\'{e}s-Aguilar, Mostovoy, Sushkov, Zhang,
  Choi, Cheong, and Drew}}]{Aguilar2009}
\bibinfo{author}{\bibfnamefont{R.}~\bibnamefont{Vald\'{e}s-Aguilar}},
  \bibinfo{author}{\bibfnamefont{M.}~\bibnamefont{Mostovoy}},
  \bibinfo{author}{\bibfnamefont{A.~B.} \bibnamefont{Sushkov}},
  \bibinfo{author}{\bibfnamefont{C.~L.} \bibnamefont{Zhang}},
  \bibinfo{author}{\bibfnamefont{Y.~J.} \bibnamefont{Choi}},
  \bibinfo{author}{\bibfnamefont{S.~W.} \bibnamefont{Cheong}},
  \bibnamefont{and} \bibinfo{author}{\bibfnamefont{H.~D.} \bibnamefont{Drew}},
  \bibinfo{journal}{Phys. Rev. Lett.} \textbf{\bibinfo{volume}{102}},
  \bibinfo{pages}{047203} (\bibinfo{year}{2009}).

\bibitem[{\citenamefont{Kenzelmann et~al.}(2005)\citenamefont{Kenzelmann,
  Harris, Jonas, Broholm, Schefer, Kim, Zhang, Cheong, Vajk, and
  Lynn}}]{Kenzelmann2005}
\bibinfo{author}{\bibfnamefont{M.}~\bibnamefont{Kenzelmann}},
  \bibinfo{author}{\bibfnamefont{A.~B.} \bibnamefont{Harris}},
  \bibinfo{author}{\bibfnamefont{S.}~\bibnamefont{Jonas}},
  \bibinfo{author}{\bibfnamefont{C.}~\bibnamefont{Broholm}},
  \bibinfo{author}{\bibfnamefont{J.}~\bibnamefont{Schefer}},
  \bibinfo{author}{\bibfnamefont{S.~B.} \bibnamefont{Kim}},
  \bibinfo{author}{\bibfnamefont{C.~L.} \bibnamefont{Zhang}},
  \bibinfo{author}{\bibfnamefont{S.~W.} \bibnamefont{Cheong}},
  \bibinfo{author}{\bibfnamefont{O.~P.} \bibnamefont{Vajk}}, \bibnamefont{and}
  \bibinfo{author}{\bibfnamefont{J.~W.} \bibnamefont{Lynn}},
  \bibinfo{journal}{Phys. Rev. Lett.} \textbf{\bibinfo{volume}{95}},
  \bibinfo{pages}{087206} (\bibinfo{year}{2005}).

\bibitem[{\citenamefont{Kimura et~al.}(2003)\citenamefont{Kimura, Goto,
  Shintani, Ishizaka, Arima, and Tokura}}]{Kimura2003a}
\bibinfo{author}{\bibfnamefont{T.}~\bibnamefont{Kimura}},
  \bibinfo{author}{\bibfnamefont{T.}~\bibnamefont{Goto}},
  \bibinfo{author}{\bibfnamefont{H.}~\bibnamefont{Shintani}},
  \bibinfo{author}{\bibfnamefont{K.}~\bibnamefont{Ishizaka}},
  \bibinfo{author}{\bibfnamefont{T.}~\bibnamefont{Arima}}, \bibnamefont{and}
  \bibinfo{author}{\bibfnamefont{Y.}~\bibnamefont{Tokura}},
  \bibinfo{journal}{Nature} \textbf{\bibinfo{volume}{426}}, \bibinfo{pages}{55}
  (\bibinfo{year}{2003}).

\bibitem[{\citenamefont{Wilkins et~al.}(2009)\citenamefont{Wilkins, Forrest,
  Beale, Bland, Walker, Mannix, Yakhou, Prabhakaran, Boothroyd, Hill
  et~al.}}]{Wilkins2009}
\bibinfo{author}{\bibfnamefont{S.~B.} \bibnamefont{Wilkins}},
  \bibinfo{author}{\bibfnamefont{T.~R.} \bibnamefont{Forrest}},
  \bibinfo{author}{\bibfnamefont{T.~A.~W.} \bibnamefont{Beale}},
  \bibinfo{author}{\bibfnamefont{S.~R.} \bibnamefont{Bland}},
  \bibinfo{author}{\bibfnamefont{H.~C.} \bibnamefont{Walker}},
  \bibinfo{author}{\bibfnamefont{D.}~\bibnamefont{Mannix}},
  \bibinfo{author}{\bibfnamefont{F.}~\bibnamefont{Yakhou}},
  \bibinfo{author}{\bibfnamefont{D.}~\bibnamefont{Prabhakaran}},
  \bibinfo{author}{\bibfnamefont{A.~T.} \bibnamefont{Boothroyd}},
  \bibinfo{author}{\bibfnamefont{J.~P.} \bibnamefont{Hill}},
  \bibnamefont{et~al.}, \bibinfo{journal}{Phys. Rev. Lett.}
  \textbf{\bibinfo{volume}{103}}, \bibinfo{pages}{207602}
  (\bibinfo{year}{2009}).

\bibitem[{\citenamefont{Aliouane et~al.}(2009)\citenamefont{Aliouane, Schmalzl,
  Senff, Maljuk, Prokes, Braden, and Argyriou}}]{Aliouane2009}
\bibinfo{author}{\bibfnamefont{N.}~\bibnamefont{Aliouane}},
  \bibinfo{author}{\bibfnamefont{K.}~\bibnamefont{Schmalzl}},
  \bibinfo{author}{\bibfnamefont{D.}~\bibnamefont{Senff}},
  \bibinfo{author}{\bibfnamefont{A.}~\bibnamefont{Maljuk}},
  \bibinfo{author}{\bibfnamefont{K.}~\bibnamefont{Prokes}},
  \bibinfo{author}{\bibfnamefont{M.}~\bibnamefont{Braden}}, \bibnamefont{and}
  \bibinfo{author}{\bibfnamefont{D.~N.} \bibnamefont{Argyriou}},
  \bibinfo{journal}{Phys. Rev. Lett.} \textbf{\bibinfo{volume}{102}},
  \bibinfo{pages}{207205} (\bibinfo{year}{2009}).

\bibitem[{\citenamefont{Yamasaki et~al.}(2007)\citenamefont{Yamasaki, Sagayama,
  Goto, Matsuura, Hirota, Arima, and Tokura}}]{Yamasaki2007}
\bibinfo{author}{\bibfnamefont{Y.}~\bibnamefont{Yamasaki}},
  \bibinfo{author}{\bibfnamefont{H.}~\bibnamefont{Sagayama}},
  \bibinfo{author}{\bibfnamefont{T.}~\bibnamefont{Goto}},
  \bibinfo{author}{\bibfnamefont{M.}~\bibnamefont{Matsuura}},
  \bibinfo{author}{\bibfnamefont{K.}~\bibnamefont{Hirota}},
  \bibinfo{author}{\bibfnamefont{T.}~\bibnamefont{Arima}}, \bibnamefont{and}
  \bibinfo{author}{\bibfnamefont{Y.}~\bibnamefont{Tokura}},
  \bibinfo{journal}{Phys. Rev. Lett.} \textbf{\bibinfo{volume}{98}},
  \bibinfo{pages}{147204} (\bibinfo{year}{2007}).

\bibitem[{\citenamefont{Kajimoto et~al.}(2009)\citenamefont{Kajimoto, Sagayama,
  Sasai, Fukuda, Tsutsui, Arima, Hirota, Mitsui, Yoshizawa, Baron
  et~al.}}]{Kajimoto2009}
\bibinfo{author}{\bibfnamefont{R.}~\bibnamefont{Kajimoto}},
  \bibinfo{author}{\bibfnamefont{H.}~\bibnamefont{Sagayama}},
  \bibinfo{author}{\bibfnamefont{K.}~\bibnamefont{Sasai}},
  \bibinfo{author}{\bibfnamefont{T.}~\bibnamefont{Fukuda}},
  \bibinfo{author}{\bibfnamefont{S.}~\bibnamefont{Tsutsui}},
  \bibinfo{author}{\bibfnamefont{T.}~\bibnamefont{Arima}},
  \bibinfo{author}{\bibfnamefont{K.}~\bibnamefont{Hirota}},
  \bibinfo{author}{\bibfnamefont{Y.}~\bibnamefont{Mitsui}},
  \bibinfo{author}{\bibfnamefont{H.}~\bibnamefont{Yoshizawa}},
  \bibinfo{author}{\bibfnamefont{A.~Q.~R.} \bibnamefont{Baron}},
  \bibnamefont{et~al.}, \bibinfo{journal}{Phys. Rev. Lett.}
  \textbf{\bibinfo{volume}{102}}, \bibinfo{pages}{247602}
  (\bibinfo{year}{2009}).

\bibitem[{\citenamefont{Malashevich and Vanderbilt}(2008)}]{Malashevich2008}
\bibinfo{author}{\bibfnamefont{A.}~\bibnamefont{Malashevich}} \bibnamefont{and}
  \bibinfo{author}{\bibfnamefont{D.}~\bibnamefont{Vanderbilt}},
  \bibinfo{journal}{Phys. Rev. Lett.} \textbf{\bibinfo{volume}{101}},
  \bibinfo{pages}{037210} (\bibinfo{year}{2008}).

\bibitem[{\citenamefont{Pimenov et~al.}(2006)\citenamefont{Pimenov, Mukhin,
  Ivanov, Travkin, Balbashov, and Loidl}}]{Pimenov2006a}
\bibinfo{author}{\bibfnamefont{A.}~\bibnamefont{Pimenov}},
  \bibinfo{author}{\bibfnamefont{A.~A.} \bibnamefont{Mukhin}},
  \bibinfo{author}{\bibfnamefont{V.~Y.} \bibnamefont{Ivanov}},
  \bibinfo{author}{\bibfnamefont{V.~D.} \bibnamefont{Travkin}},
  \bibinfo{author}{\bibfnamefont{A.~M.} \bibnamefont{Balbashov}},
  \bibnamefont{and} \bibinfo{author}{\bibfnamefont{A.}~\bibnamefont{Loidl}},
  \bibinfo{journal}{Nat. Phys.} \textbf{\bibinfo{volume}{2}},
  \bibinfo{pages}{97} (\bibinfo{year}{2006}).

\bibitem[{\citenamefont{Senff et~al.}(2007)\citenamefont{Senff, Link, Hradil,
  Hiess, Regnault, Sidis, Aliouane, Argyriou, and Braden}}]{Senff2007}
\bibinfo{author}{\bibfnamefont{D.}~\bibnamefont{Senff}},
  \bibinfo{author}{\bibfnamefont{P.}~\bibnamefont{Link}},
  \bibinfo{author}{\bibfnamefont{K.}~\bibnamefont{Hradil}},
  \bibinfo{author}{\bibfnamefont{A.}~\bibnamefont{Hiess}},
  \bibinfo{author}{\bibfnamefont{L.~P.} \bibnamefont{Regnault}},
  \bibinfo{author}{\bibfnamefont{Y.}~\bibnamefont{Sidis}},
  \bibinfo{author}{\bibfnamefont{N.}~\bibnamefont{Aliouane}},
  \bibinfo{author}{\bibfnamefont{D.~N.} \bibnamefont{Argyriou}},
  \bibnamefont{and} \bibinfo{author}{\bibfnamefont{M.}~\bibnamefont{Braden}},
  \bibinfo{journal}{Phys. Rev. Lett.} \textbf{\bibinfo{volume}{98}},
  \bibinfo{pages}{137206} (\bibinfo{year}{2007}).

\bibitem[{\citenamefont{Senff et~al.}(2008)\citenamefont{Senff, Aliouane,
  Argyriou, Hiess, Regnault, Link, Hradil, Sidis, and Braden}}]{Senff2008a}
\bibinfo{author}{\bibfnamefont{D.}~\bibnamefont{Senff}},
  \bibinfo{author}{\bibfnamefont{N.}~\bibnamefont{Aliouane}},
  \bibinfo{author}{\bibfnamefont{D.~N.} \bibnamefont{Argyriou}},
  \bibinfo{author}{\bibfnamefont{A.}~\bibnamefont{Hiess}},
  \bibinfo{author}{\bibfnamefont{L.~P.} \bibnamefont{Regnault}},
  \bibinfo{author}{\bibfnamefont{P.}~\bibnamefont{Link}},
  \bibinfo{author}{\bibfnamefont{K.}~\bibnamefont{Hradil}},
  \bibinfo{author}{\bibfnamefont{Y.}~\bibnamefont{Sidis}}, \bibnamefont{and}
  \bibinfo{author}{\bibfnamefont{M.}~\bibnamefont{Braden}},
  \bibinfo{journal}{Journal Of Physics-Condensed Matter}
  \textbf{\bibinfo{volume}{20}}, \bibinfo{pages}{434212}
  (\bibinfo{year}{2008}).

\bibitem[{\citenamefont{Rovillain et~al.}(2010)\citenamefont{Rovillain,
  Cazayous, Gallais, Sacuto, Measson, and Sakata}}]{Rovillain2010}
\bibinfo{author}{\bibfnamefont{P.}~\bibnamefont{Rovillain}},
  \bibinfo{author}{\bibfnamefont{M.}~\bibnamefont{Cazayous}},
  \bibinfo{author}{\bibfnamefont{Y.}~\bibnamefont{Gallais}},
  \bibinfo{author}{\bibfnamefont{A.}~\bibnamefont{Sacuto}},
  \bibinfo{author}{\bibfnamefont{M.-A.} \bibnamefont{Measson}},
  \bibnamefont{and} \bibinfo{author}{\bibfnamefont{H.}~\bibnamefont{Sakata}},
  \bibinfo{journal}{Phys. Rev. B} \textbf{\bibinfo{volume}{81}},
  \bibinfo{pages}{054428} (\bibinfo{year}{2010}).

\bibitem[{\citenamefont{Takahashi et~al.}(2008)\citenamefont{Takahashi, Kida,
  Yamasaki, Fujioka, Arima, Shimano, Miyahara, Mochizuki, Furukawa, and
  Tokura}}]{Takahashi2008}
\bibinfo{author}{\bibfnamefont{Y.}~\bibnamefont{Takahashi}},
  \bibinfo{author}{\bibfnamefont{N.}~\bibnamefont{Kida}},
  \bibinfo{author}{\bibfnamefont{Y.}~\bibnamefont{Yamasaki}},
  \bibinfo{author}{\bibfnamefont{J.}~\bibnamefont{Fujioka}},
  \bibinfo{author}{\bibfnamefont{T.}~\bibnamefont{Arima}},
  \bibinfo{author}{\bibfnamefont{R.}~\bibnamefont{Shimano}},
  \bibinfo{author}{\bibfnamefont{S.}~\bibnamefont{Miyahara}},
  \bibinfo{author}{\bibfnamefont{M.}~\bibnamefont{Mochizuki}},
  \bibinfo{author}{\bibfnamefont{N.}~\bibnamefont{Furukawa}}, \bibnamefont{and}
  \bibinfo{author}{\bibfnamefont{Y.}~\bibnamefont{Tokura}},
  \bibinfo{journal}{Phys. Rev. Lett.} \textbf{\bibinfo{volume}{101}},
  \bibinfo{pages}{187201} (\bibinfo{year}{2008}).

\bibitem[{\citenamefont{Pimenov et~al.}(2009)\citenamefont{Pimenov, Shuvaev,
  Loidl, Schrettle, Mukhin, Travkin, Ivanov, and Balbashov}}]{Pimenov2009}
\bibinfo{author}{\bibfnamefont{A.}~\bibnamefont{Pimenov}},
  \bibinfo{author}{\bibfnamefont{A.}~\bibnamefont{Shuvaev}},
  \bibinfo{author}{\bibfnamefont{A.}~\bibnamefont{Loidl}},
  \bibinfo{author}{\bibfnamefont{F.}~\bibnamefont{Schrettle}},
  \bibinfo{author}{\bibfnamefont{A.~A.} \bibnamefont{Mukhin}},
  \bibinfo{author}{\bibfnamefont{V.~D.} \bibnamefont{Travkin}},
  \bibinfo{author}{\bibfnamefont{V.~Y.} \bibnamefont{Ivanov}},
  \bibnamefont{and} \bibinfo{author}{\bibfnamefont{A.~M.}
  \bibnamefont{Balbashov}}, \bibinfo{journal}{Phys. Rev. Lett.}
  \textbf{\bibinfo{volume}{102}}, \bibinfo{pages}{107203}
  (\bibinfo{year}{2009}).

\bibitem[{\citenamefont{Schmidt et~al.}(2009)\citenamefont{Schmidt, Kant,
  Rudolf, Mayr, Mukhin, Balbashov, Deisenhofer, and Loidl}}]{Schmidt2009}
\bibinfo{author}{\bibfnamefont{M.}~\bibnamefont{Schmidt}},
  \bibinfo{author}{\bibfnamefont{C.}~\bibnamefont{Kant}},
  \bibinfo{author}{\bibfnamefont{T.}~\bibnamefont{Rudolf}},
  \bibinfo{author}{\bibfnamefont{F.}~\bibnamefont{Mayr}},
  \bibinfo{author}{\bibfnamefont{A.~A.} \bibnamefont{Mukhin}},
  \bibinfo{author}{\bibfnamefont{A.~M.} \bibnamefont{Balbashov}},
  \bibinfo{author}{\bibfnamefont{J.}~\bibnamefont{Deisenhofer}},
  \bibnamefont{and} \bibinfo{author}{\bibfnamefont{A.}~\bibnamefont{Loidl}},
  \bibinfo{journal}{European Physical Journal B} \textbf{\bibinfo{volume}{71}},
  \bibinfo{pages}{411} (\bibinfo{year}{2009}).

\bibitem[{\citenamefont{Sakata et~al.}(2007)\citenamefont{Sakata, Hosokawa, and
  Kato}}]{Sakata2007}
\bibinfo{author}{\bibfnamefont{H.}~\bibnamefont{Sakata}},
  \bibinfo{author}{\bibfnamefont{K.}~\bibnamefont{Hosokawa}}, \bibnamefont{and}
  \bibinfo{author}{\bibfnamefont{T.}~\bibnamefont{Kato}},
  \bibinfo{journal}{Int. J. Mod. Phys. B} \textbf{\bibinfo{volume}{21}},
  \bibinfo{pages}{3425} (\bibinfo{year}{2007}).

\bibitem[{\citenamefont{Homes et~al.}(1993)\citenamefont{Homes, Reedyk,
  Crandles, and Timusk}}]{Homes1993}
\bibinfo{author}{\bibfnamefont{C.~C.} \bibnamefont{Homes}},
  \bibinfo{author}{\bibfnamefont{M.}~\bibnamefont{Reedyk}},
  \bibinfo{author}{\bibfnamefont{D.~A.} \bibnamefont{Crandles}},
  \bibnamefont{and} \bibinfo{author}{\bibfnamefont{T.}~\bibnamefont{Timusk}},
  \bibinfo{journal}{Appl. Opt.} \textbf{\bibinfo{volume}{32}},
  \bibinfo{pages}{2976} (\bibinfo{year}{1993}).

\bibitem[{\citenamefont{Duarte et~al.}(1987)\citenamefont{Duarte, Sanjurjo, and
  Katiyar}}]{Duarte1987}
\bibinfo{author}{\bibfnamefont{J.~L.} \bibnamefont{Duarte}},
  \bibinfo{author}{\bibfnamefont{J.~A.} \bibnamefont{Sanjurjo}},
  \bibnamefont{and} \bibinfo{author}{\bibfnamefont{R.~S.}
  \bibnamefont{Katiyar}}, \bibinfo{journal}{Phys. Rev. B}
  \textbf{\bibinfo{volume}{36}}, \bibinfo{pages}{3368} (\bibinfo{year}{1987}).

\bibitem[{\citenamefont{Schleck et~al.}(2010)\citenamefont{Schleck, Nahas,
  Lobo, Varignon, Lepetit, Nelson, and Moreira}}]{Schleck2009}
\bibinfo{author}{\bibfnamefont{R.}~\bibnamefont{Schleck}},
  \bibinfo{author}{\bibfnamefont{Y.}~\bibnamefont{Nahas}},
  \bibinfo{author}{\bibfnamefont{R.~P. S.~M.} \bibnamefont{Lobo}},
  \bibinfo{author}{\bibfnamefont{J.}~\bibnamefont{Varignon}},
  \bibinfo{author}{\bibfnamefont{M.~B.} \bibnamefont{Lepetit}},
  \bibinfo{author}{\bibfnamefont{C.~S.} \bibnamefont{Nelson}},
  \bibnamefont{and} \bibinfo{author}{\bibfnamefont{R.~L.}
  \bibnamefont{Moreira}}, \bibinfo{journal}{Phys. Rev. B}
  \textbf{\bibinfo{volume}{XX}}, \bibinfo{pages}{PPPP} (\bibinfo{year}{2010}),
  \eprint{arXiv:0910.3137}.

\bibitem[{\citenamefont{Laverdiere et~al.}(2006)\citenamefont{Laverdiere,
  Jandl, Mukhin, Ivanov, Ivanov, and Iliev}}]{Laverdiere2006}
\bibinfo{author}{\bibfnamefont{J.}~\bibnamefont{Laverdiere}},
  \bibinfo{author}{\bibfnamefont{S.}~\bibnamefont{Jandl}},
  \bibinfo{author}{\bibfnamefont{A.~A.} \bibnamefont{Mukhin}},
  \bibinfo{author}{\bibfnamefont{V.~Y.} \bibnamefont{Ivanov}},
  \bibinfo{author}{\bibfnamefont{V.~G.} \bibnamefont{Ivanov}},
  \bibnamefont{and} \bibinfo{author}{\bibfnamefont{M.~N.} \bibnamefont{Iliev}},
  \bibinfo{journal}{Phys. Rev. B} \textbf{\bibinfo{volume}{74}},
  \bibinfo{pages}{179902} (\bibinfo{year}{2006}).

\bibitem[{\citenamefont{Xiang et~al.}(2008)\citenamefont{Xiang, Wei, Whangbo,
  and Da~Silva}}]{Xiang2008}
\bibinfo{author}{\bibfnamefont{H.~J.} \bibnamefont{Xiang}},
  \bibinfo{author}{\bibfnamefont{S.~H.} \bibnamefont{Wei}},
  \bibinfo{author}{\bibfnamefont{M.~H.} \bibnamefont{Whangbo}},
  \bibnamefont{and} \bibinfo{author}{\bibfnamefont{J.~L.~F.}
  \bibnamefont{Da~Silva}}, \bibinfo{journal}{Phys. Rev. Lett.}
  \textbf{\bibinfo{volume}{101}}, \bibinfo{pages}{037209}
  (\bibinfo{year}{2008}).

\bibitem[{\citenamefont{Goto et~al.}(2004)\citenamefont{Goto, Kimura, Lawes,
  Ramirez, and Tokura}}]{Goto2004}
\bibinfo{author}{\bibfnamefont{T.}~\bibnamefont{Goto}},
  \bibinfo{author}{\bibfnamefont{T.}~\bibnamefont{Kimura}},
  \bibinfo{author}{\bibfnamefont{G.}~\bibnamefont{Lawes}},
  \bibinfo{author}{\bibfnamefont{A.~P.} \bibnamefont{Ramirez}},
  \bibnamefont{and} \bibinfo{author}{\bibfnamefont{Y.}~\bibnamefont{Tokura}},
  \bibinfo{journal}{Phys. Rev. Lett.} \textbf{\bibinfo{volume}{92}},
  \bibinfo{pages}{257201} (\bibinfo{year}{2004}).

\bibitem[{\citenamefont{Barath et~al.}(2008)\citenamefont{Barath, Kim, Cooper,
  Abbamonte, Fradkin, Mahns, R\"{u}bhausen, Aliouane, and
  Argyriou}}]{Barath2008}
\bibinfo{author}{\bibfnamefont{H.}~\bibnamefont{Barath}},
  \bibinfo{author}{\bibfnamefont{M.}~\bibnamefont{Kim}},
  \bibinfo{author}{\bibfnamefont{S.~L.} \bibnamefont{Cooper}},
  \bibinfo{author}{\bibfnamefont{P.}~\bibnamefont{Abbamonte}},
  \bibinfo{author}{\bibfnamefont{E.}~\bibnamefont{Fradkin}},
  \bibinfo{author}{\bibfnamefont{I.}~\bibnamefont{Mahns}},
  \bibinfo{author}{\bibfnamefont{M.}~\bibnamefont{R\"{u}bhausen}},
  \bibinfo{author}{\bibfnamefont{N.}~\bibnamefont{Aliouane}}, \bibnamefont{and}
  \bibinfo{author}{\bibfnamefont{D.~N.} \bibnamefont{Argyriou}},
  \bibinfo{journal}{Phys. Rev. B} \textbf{\bibinfo{volume}{78}},
  \bibinfo{pages}{134407} (\bibinfo{year}{2008}).

\end{thebibliography}

\end{document}